\documentclass[prl,twocolumn,showpacs,floatfix]{revtex4}%
\usepackage{amsfonts}
\usepackage{amsmath}
\usepackage{amssymb}
\usepackage{graphicx}
\usepackage{epsfig}%
\begin{document}
\title{Feedback cooling of a single trapped ion}
\author{Pavel Bushev$^*$, Daniel Rotter, Alex Wilson, Fran\c{c}ois Dubin, Christoph
Becher$^\S$, J{\"u}rgen Eschner$^\dag$, and Rainer
Blatt$^{\ddagger}$} \affiliation{Institute for Experimental
Physics, University of Innsbruck, Technikerstr.\ 25, A-6020
Innsbruck, Austria}
\author{Viktor Steixner$^{\ddagger}$, Peter Rabl$^{\ddagger}$, and Peter Zoller$^{\ddagger}$ }
\affiliation{Institute for Theoretical Physics, University of Innsbruck,
Technikerstr.\ 25, A-6020 Innsbruck, Austria\\ \ddag: Institute for
Quantum Optics and Quantum Information of the Austrian Academy of
Sciences, 6020 Innsbruck, Austria }
\date{\today }

\begin{abstract}
Based on a real-time measurement of the motion of a single ion in
a Paul trap, we demonstrate its electro-mechanical cooling below
the Doppler limit by homodyne feedback control (cold damping). The
feedback cooling results are well described by a model based on a
quantum mechanical Master Equation.

\end{abstract}

\pacs{3.65.Ta,  42.50.Lc  32.80.Pj, 42.50.Ct, 42.50.Vk, 32.80.Lg  }
\maketitle

Quantum optics, and more recently mesoscopic condensed matter
physics, have taken a leading role in realizing individual quantum
systems, which can be monitored continuously in quantum limited
measurements, and at the same time can be controlled by external
fields on time scales fast in comparison with the system
evolution. Examples include cold trapped ions and atoms
\cite{LeibfriedRMP03}, cavity QED \cite{Keller2004, McKeever2004,
Kuhn2002, Orozco2004} and nanomechanical systems
\cite{hopkins2003}. This setting opens the possibility of
manipulating individual quantum systems by feedback, a problem
which is not only of a fundamental interest in quantum mechanics,
but also promises a new route to generating interesting quantum
states in the laboratory. First experimental efforts to realize
quantum feedback have been reported only recently. While not all
of them may qualify as quantum feedback in a strict sense,
feedback has been applied to various quantum systems
\cite{Gabrielse2003, Steck2004, Rempe2002, Morrow2002, Orozco2004,
Geremia2004}. On the theory side, this has motivated during the
last decade the development of a \emph{quantum feedback theory}
\cite{Wiseman1993, Vitali2002}, where the basic ingredients are
the interplay between quantum dynamics and the back-action of the
measurement on the system evolution. In this letter we report a
first experiment to demonstrate quantum feedback control, i.e.
quantum feedback cooling, of a single trapped ion by monitoring
the fluorescence of the laser driven ion in front of a mirror. We
establish a continuous measurement of the position of the ion
which allows us to act back in a feedback loop demonstrating
``cold damping'' \cite{Milatz1953, Cohadon1999}. We will show that
quantum control theory based on a quantum optical modelling of the
system dynamics and continuous measurement theory of
photodetection provides a quantitative understanding of the
experimental results.

\begin{figure}[b]
\epsfig{file=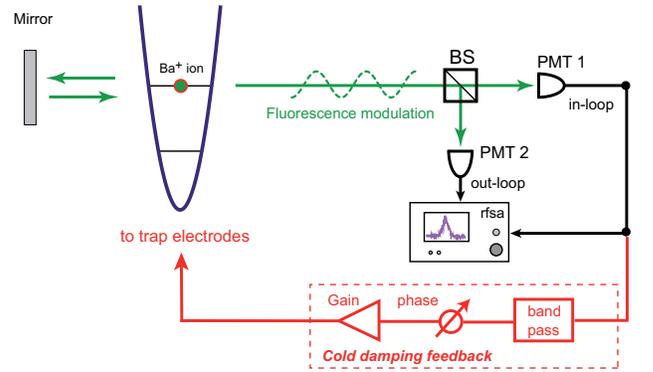, width=0.95\hsize}
\caption{A single $^{138}$Ba$^{+}$ ion in a Paul trap (parabola)
is laser-excited and -cooled on its S$_{1/2}$ to P$_{1/2}$
transition at 493~nm. A retro-reflecting mirror 25~cm away from
the trap and a lens (not shown) focus back the fluorescence onto
the ion. The resulting interference fringes with up to $73\%$
contrast are observed by a photomultiplier (PMT 1). The ion's
oscillation in the trap creates an intensity modulation of the PMT
signal which is observed as a sideband on a spectrum analyser
(rfsa) \cite{Bushev2004}. For feedback cooling, the sideband
signal is filtered, phase-shifted, and applied to the ion as a
voltage on a trap electrode.}%
\label{setup_feedback}%
\end{figure}

We study a single $^{138}$Ba$^{+}$ ion in a miniature Paul trap
which is continuously laser-excited and laser-cooled to the
Doppler limit on its S$_{1/2}$ to P$_{1/2}$ transition at 493~nm,
as outlined in Fig.~1. The ion is driven by a laser near the
atomic resonance, and the scattered light is emitted both into the
radiation modes reflected by the mirror, as well as the other
(background) modes of the quantized light field \cite{Dorner2002}.
Light scattered into the mirror modes can either reach the
photodetector directly, or after reflection from the mirror. From
the resulting interference the motion of the ion (its projection
onto the ion-mirror axis) is detected as a vibrational sideband in
the fluctuation spectrum of the photon counting signal
\cite{Bushev2004}. Of the three sidebands at about
($1$,$1.2$,$2.3$)~MHz, corresponding to the three axes of
vibration, we observe the one at $\nu=1$~MHz. It has a width
$\Gamma\approx400$~Hz and is superimposed on the background shot
noise generated by the photon counting process.

Our goal is to continuously read the position of the ion, and feed
back a damping force proportional to the momentum to achieve
feedback cooling. For a weakly driven atom the emitted light is
dominantly elastic scattering at the laser frequency, and the
information on the motion of the ion is encoded in the sidebands
of the scattered light, displaced by the trap frequency $\nu$. For
an ion trapped in the Lamb-Dicke regime ($\eta \sqrt{N} \ll 1$,
with the Lamb Dicke parameter $\eta = 2\pi a_0 / \lambda \sim
0.07$, $a_0$ the r.m.s. size of the trap ground state, $N$ the
mean excitation number of the motional oscillator) the motional
sidebands are suppressed by $\eta$ relative to the elastic
component at the laser frequency. Thus the light reaching the
detector will be the sum of the elastic component and weak
sidebands \cite{Bushev2004}, a situation reminiscent of homodyne
detection, where a strong oscillator beats with the signal field
to provide a homodyne current at the detector. This physical
picture allows us to formulate the continuous readout of the
position of the ion as well as the quantum feedback cooling in the
well-developed language of homodyne detection and quantum
feedback.

The homodyne current at the photodetector (see Fig.~1) with the (large) signal
from the elastic light scattering subtracted has the form
\begin{equation}
I_{c}(t)=\gamma\eta\left\langle \hat{z}\right\rangle _{c}(t)+\sqrt
{\frac{\gamma}{2}}\xi(t)~. \label{eq:photocurrent}%
\end{equation}
The first term is proportional to the conditioned expectation
value of the position of the trapped ion, $\left\langle
\hat{z}\right\rangle _{c}(t)$, and the second term is a shot noise
contribution with Gaussian white noise $\xi(t)$. We have defined
$\hat{z}=a+a^{\dag}\equiv z/a_{0}$ with $a$ ($a^{\dag}$)
destruction (creation) operator of the harmonic oscillator, and we
have assumed that the trap center is placed at a distance
$L=n\lambda /2+\lambda/8$ ($n$ integer) from the mirror,
corresponding to a point of maximum slope of the standing wave
intensity of the mirror mode. The current $I_{c}(t)$ scales with
$\gamma \propto\epsilon$, which is the light scattering rate into
the solid angle ($4\pi\epsilon$) of the mirror mode induced by the
laser. The expectation value $\left\langle \cdot\right\rangle _{c}
\equiv\operatorname{Tr}\{\cdot\,\rho_{c}(t)\}$ is defined with
respect to a conditioned density operator $\rho_{c}(t)$, which
reflects our knowledge of the motional state of the ion for the
given history of the photocurrent. According to the theory of
homodyne detection, $\rho_{c}(t)$ obeys the Ito stochastic
differential equation \cite{GZ}
\begin{align}
d\rho_{c}(t)=  &  -\mathrm{i}\nu\lbrack a^{\dag}a,\rho_{c}(t)]dt+\mathcal{L}%
_{0}\rho_{c}(t)dt\nonumber\\
&  +\sqrt{2\gamma\eta^{2}}~\mathcal{H}\rho_{c}(t)dW(t)~, \label{eq:Ito}%
\end{align}
where $\mathcal{H}\rho_{c}(t)=(\hat{z}\rho_{c}(t)+\rho_{c}(t)\hat
{z}-2\left\langle \hat{z}\right\rangle _{c}(t)\rho_{c}(t))\,.$ The
first line determines the unobserved evolution of the ion,
including the harmonic motion in the trap with frequency $\nu$ and
the dissipative dynamics, $\mathcal{L}_{0}$, due to photon
scattering. The latter is given by
\begin{equation}
\mathcal{L}_{0}\rho=\Gamma(N+1)\mathcal{D}[a]\rho+\Gamma
N\mathcal{D}[a^{\dag }]\rho~,
\end{equation}
where we have defined the superoperator $\mathcal{D}[c]\rho\equiv
c\rho c^{\dag}-(c^{\dag}c\rho+\rho c^{\dag}c)/2$. The laser
cooling rate $\Gamma$ and the steady state occupation number,
$N=\langle a^{\dag}a\rangle$, can either be estimated from the
motional sidebands or deduced from the cooling laser parameters
\cite{LeibfriedRMP03}. In the present experiment, $N\approx17$
corresponds to the Doppler limit. The last term of
Eq.~\eqref{eq:Ito} is proportional to the Wiener increment
$dW(t)\equiv\xi(t)dt$ and corresponds to an update of the
observers knowledge about the system according to a certain
measurement result $I_{c}(t)$. In summary,
Eq.~\eqref{eq:photocurrent} demonstrates that observation of the
sidebands of the light scattered into the mirror mode provides us
with information of the position of the ion, while the system
density matrix is updated according to Eq.~\eqref{eq:Ito}. This is
the basis for describing feedback control of the ion, as shown in
the following.

For feedback cooling, the vibrational sideband is extracted with a
bandpass filter of bandwidth $B=30$~kHz, shifted by $(-\pi/2)$,
amplified, and the resulting output voltage is applied to an
electrode which is close to the trap inside the vacuum. Thereby we
create a driving force which is proportional to (and opposed to)
the instantaneous velocity of the ion and which thus adds to the
damping of its vibration. The overall gain of the feedback loop
depends on many factors such as the fringe contrast of the
interference, the PMT characteristics, etc. It is varied
electronically by setting the amplification $G$ of the final
amplifier in the loop. We can also set the phase to other values
than the optimum of $(-\pi/2)$, which will be used in order to
compare experiment and theory.

To analyze the result of the feedback we look at the changes in
the sideband spectrum. The modified spectra require careful
interpretation. The spectrum observed inside the feedback loop
(``in-loop'' or PMT1 in Fig.~\ref{setup_feedback}) shows not only
the motion of the ion, but the sum of the motion and the shot
noise. As the feedback correlates these fluctuations, a reduction
of the signal below the shot noise level may occur, similar in
appearance to signatures of squeezed light. This effect is known
as ``squashing'' \cite{Buchler1999}, or ``anti-correlated state of
light'' in an opto-electronic loop \cite{Masalov1994}, and it does
not constitute a quantum mechanical squeezing of the fluctuations
\cite{Haus1986, Shapiro1987}. The effect on the motion can only be
reliably detected by splitting the optical signal before it is
measured and recording it outside the feedback loop with a second
PMT (``out-loop'' in Fig.~\ref{setup_feedback}), whose shot noise
is not correlated with the motion.

\begin{figure}[ptb]
\epsfig{file=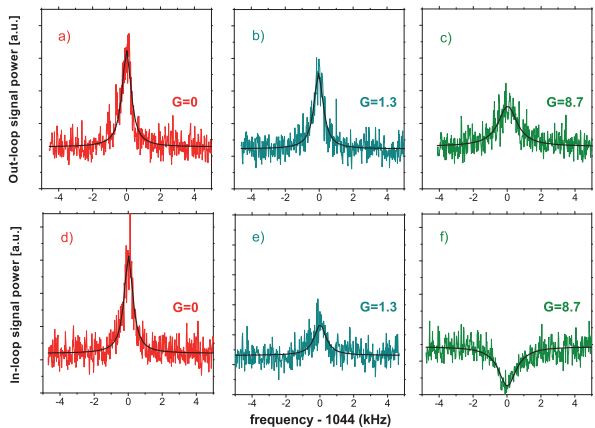, width=0.90\hsize}
\caption{Feedback cooling spectra. The vibrational sideband around
$\nu=1$~MHz is shown on top of the spectrally flat shot noise
background, to which the spectra are normalized. The upper curves
(a)-(c) are measured outside the feedback loop, while the lower
curves (d)-(f) are the in-loop results. Spectra (a) and (d) are
for laser cooling only, the other curves are recorded with
feedback at the indicated gain values. The feedback phase is set
to $(-\pi/2)$. The gain values indicate the settings of the
final amplifier in the feedback loop.}%
\label{inloop_outloop_sideband}%
\end{figure}

In Fig.~\ref{inloop_outloop_sideband} we show spectra recorded
with the spectrum analyzer, and measured outside and inside the
feedback loop. The first curve of each row, showing the largest
sideband, is the one without feedback (gain $G=0$). The other two
curves are recorded with the loop closed (gain values $G=1.3$ and
$8.7$). The sub shot noise fluctuations inside the loop, when the
ion is driven to move in anti-phase with the shot noise, are
clearly visible in Fig.~\ref{inloop_outloop_sideband}(f). The main
cooling results are curves (b) and (c) which show the motional
sideband reduced in size and broadened, indicating a reduced
energy (proportional to the area under the curve) and a higher
damping rate (the width). From case (b) to (c) the area increases,
as the injected and amplified shot noise overcompensates the
increased damping. As shown below, in our model the incorporation
of quantum feedback competing with laser cooling predicts such
behavior, i.e., the existence of an optimal gain for maximal
cooling (for a detailed description cf. \cite{Steixner2005}).

We model the effect of the feedback force acting on the ion by
extending Eq.~\eqref{eq:Ito} with the feedback contribution,
\begin{equation}
\lbrack d\rho_{c}(t)]_{\mathrm{fb}}=-\mathrm{i}\tilde{G}I_{\mathrm{fb}%
,c}(t-\tau)[\hat{z},\rho_{c}(t)], \label{eq:hfb}%
\end{equation}
where $I_{\mathrm{fb},c}(t)$ denotes the measured current after
the feedback circuit. The latter is proportional to the voltage
applied to the trap electrodes. All conversion factors between the
feedback current and the actual force applied on the ion are
included in the overall gain $\tilde G$. The time delay $\tau$ in
the feedback loop preserves causality and is small compared to the
fastest timescale $\nu^{-1}$ of the motion of the ion which allows
us to consider the Markovian limit ($\tau\rightarrow 0^+$).

To obtain an expression for the feedback current
$I_{\mathrm{fb},c}(t)$, we change into a frame rotating with the
trap frequency $\nu$ and define the density operator,
$\mu_{c}(t)\equiv\exp(\mathrm{i}\nu a^{\dag}at)\rho
_{c}(t)\exp(-\mathrm{i}\nu a^{\dag}at)$, evolving on the (slow)
cooling timescales. This is convenient due to the large separation
between the timescale of the harmonic oscillations, $\nu^{-1}$ and
the timescale of laser and feedback cooling in our experiment. For
our experimental parameters, $\nu\gg B\gg\Gamma$, the feedback
current for a phase shift of $(-\pi/2)$ has
the form \cite{Steixner2005}%
\begin{equation}
I_{\mathrm{fb},c}(t)=\left(  \gamma\eta\left\langle \hat{p}\right\rangle
_{c}(t)+\sqrt{\frac{\gamma}{2}}\Xi(t)\right)  \cos(\nu t),\label{eq:vfb}%
\end{equation}
where $\left\langle \hat{p}\right\rangle _{c}(t)\equiv\mathrm{Tr}\{\hat{p}%
\mu_{c}(t)\}$, and $\hat{p}\equiv\mathrm{i}(a^{\dag}-a)$ is the
momentum operator conjugated to $\hat z$. The first term in this
current therefore provides damping for the motion of the ion. The
second term of Eq.~(\ref{eq:vfb}) describes the shot noise which
passes through the electronic circuit and is fed back to the ion.
The stochastic variable $\Xi(t)$ is Gaussian white noise on a
timescale given by the inverse  bandwidth $B^{-1}$, whereby
$B\gg\Gamma$ implies that it is spectrally flat on the frequency
range of the cooling dynamics.

For a full record of the photocurrent $I_c(t)$,  Eqs.
(\ref{eq:Ito}) and (\ref{eq:hfb}) determine the evolution of the
ion's motional state in presence of feedback.  As it is
impractical to keep track of the whole photocurrent in the
experiment, we derive a master equation for the density operator
averaged over all possible realisation of $I_c(t)$, $\mu(t)$.
Along the ways of the Wiseman-Milburn theory of quantum feedback
\cite{Wiseman1993}, for a phase shift of $(-\pi/2)$, we obtain the
\emph{quantum feedback master equation} \cite{Steixner2005}
\begin{equation}
\dot{\mu}=\mathcal{L}_{0}\mu-\mathrm{i}\frac{\tilde G\gamma\eta}{4}[\hat{z},\hat
{p}\mu+\mu\hat{p}]-\frac{\tilde G^{2}\gamma}{16}[\hat{z},[\hat{z},\mu
]].\label{eq:meq_final}%
\end{equation}
The second and third term are the additional contributions due to
the feedback. The part linear in $\tilde G$ induces damping of the
motion of the ion, and the term quadratic in $\tilde G$ describes
the effect of the fed back noise leading to diffusion of the
momentum. The competition between laser cooling, damping and the
injected noise leads to the characteristic behavior of the steady
state
number expectation value%

\begin{equation}
\left\langle n\right\rangle _{ss}=\frac{N+\eta\gamma\tilde{G}(2N-1)/2\Gamma
+\gamma\tilde{G}^{2}/8\Gamma}{1+2\eta\gamma\tilde{G}/\Gamma}.
\end{equation}
For small gain, damping dominates, and the energy of the ion is
decreased below the Doppler limit. For higher gain, the diffusive
term describing the noise fed back into the system overcompensates
cooling, i.e., heats the ion. Consequently, for $(-\pi/2)$
feedback phase and optimal gain conditions the steady state energy
is minimized. On the contrary, for a $\pi$ phase shift
$(-\hat{z})$ replaces $\hat{p}$ in the second term of
Eq.~(\ref{eq:meq_final}), the feedback force then merely induces a
frequency shift, $\Delta\omega=\tilde{G}\gamma\eta/2$, but no
damping. Increasing the gain then always enhances the steady state
number expectation value, i.e., the mean ion energy. We now
compare the theoretical predictions and the experimental results
for the feedback phases of physical interest, namely $(-\pi/2)$
and $\pi$.

\begin{figure}[t]
\epsfig{file=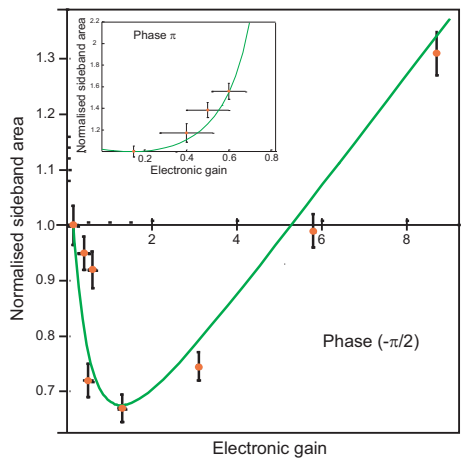, width=0.7\hsize}
\caption{Steady-state energy of the cooled oscillator: measured
sideband area, normalized to the value without feedback, versus
gain of the feedback loop, for $(-\pi/2)$ (a) and $\pi$ (b)
feedback phase. The curves are the model calculations. The gain
axis is scaled to the experimental values of the
electronic gain.}%
\label{Area_vs_Gain}%
\end{figure}

The measured ion energy as a function of the feedback electronic
gain is shown in Fig. \ref{Area_vs_Gain}. On the first side, as
expected, for a $(-\pi/2)$ feedback phase, cooling by more than
$30\%$ below the Doppler limit is achieved, while further increase
of the gain drives the shot noise and therefore heats the motion
of the ion. On the other side, a $\pi$ phase shift in the feedback
loop does not yield any damping, in such conditions the motion of
the ion is only driven. This results in an increase of the
measured sideband area (as shown in the insert of Figure
\ref{Area_vs_Gain}), as well as a shift of the sideband center
frequency (graph not shown). Both cases demonstrate good agreement
between experiment and theory. Finally, let us stress that the
optimal cooling rate is governed by the collection efficiency of
the fluorescence going into the mirror mode, $\epsilon$. In the
experiments presented above, $\epsilon\approx 1\%$ leads to a
decrease of the steady state occupation number $N$ from 17 to 12,
while $N\approx 3$ can be reached for $\epsilon\approx 15\%$,
experimentally achievable for optimal optical coupling.

To summarize, we have demonstrated real-time feedback cooling of
the motion of a single trapped ion. Electro-mechanical back-action
based on a sensitive real-time measurement of the motion of the
ion in the trap allowed us to cool one motional degree of freedom
by $30\%$ below the Doppler limit. Unlike with laser cooling, the
presented method allows to cool one of the ion's motional mode
without heating the two others, and our procedure can easily be
extended to cooling all motional modes. The cooling process is
shot-noise limited and the fraction of scattered photons recorded
to observe the motion of the ion limits the ultimate cooling at
optimum gain. The latter can yield a steady state occupation
number $N=3$ for realistic experimental conditions. In conclusion,
our feedback scheme offers a possible way to very efficiently cool
the motion of ions unsuitable for sideband cooling.

\noindent\textbf{Acknowledgements.} This work has been supported
by the Austrian Science Fund (FWF) in the project SFB15, by the
European Commission (QUEST network, HPRNCT\--2000\--00121, QUBITS
network, IST\--1999\--13021), and by the ``Institut f\"ur
Quanteninformation GmbH''. We thank S. Mancini, A. Masalov, and D.
Vitali for clarifying discussions. P.~B. thanks the group of
theoretical physics at U. Camerino, Italy, for hospitality.


\begin{thebibliography}{9999999}


\bibitem[$^*$]{1}{Present address: Laboratory of Physical Chemistry, ETH
Z{\"u}rich, Switzerland.}

\bibitem[$^\S$]{2}{Present address: Fachrichtung Technische Physik,
 Saarland University, Saarbr\"{u}cken, Germany.}

\bibitem[$^\dag$]{3}{Present address: ICFO - Institut de Ci{\`e}ncies
Fot{\`o}niques, Barcelona, Spain.}

\bibitem {LeibfriedRMP03}
D. Leibfried, R. Blatt, C. Monroe, and D. Wineland, Rev. Mod. Phys
\textbf{75}, 281 (2003), and references cited.

\bibitem {Keller2004}
M. Keller, B. Lange, K. Hayasaka, W. Lange, H. Walther, Nature \textbf{431},
1075 (2004).

\bibitem {McKeever2004}
J. McKeever, A. Boca, A. D. Boozer, R. Miller, J. R. Buck, A. Kuzmich, and H.
J. Kimble, Science \textbf{303}, 1992 (2004);
P. Maunz, T. Puppe, I. Schuster, N. Syassen, P. W. H. Pinkse, and G.
Rempe, Nature \textbf{428}, 50 (2004)

\bibitem {Kuhn2002}
T. Legero, T. Wilk, M. Hennrich, G. Rempe, and A. Kuhn, Phys. Rev. Lett.
\textbf{93}, 070503 (2004).

\bibitem {Orozco2004}
J. E. Reiner, W. P. Smith, L. A. Orozco, H. M. Wiseman, and Jay Gambetta,
Phys. Rev. A. \textbf{70}, 023819 (2004).

\bibitem{hopkins2003}A. Hopkins, K. Jacobs, S. Habib, and K. Schwab, Phys. Rev. B
\textbf{68}, 235328 (2003)

\bibitem {Steck2004}
D. A. Steck, K. Jacobs, H. Mabuchi, T. Bhattacharya, and S. Habib,
Phys. Rev. Lett \textbf{92}, 223004 (2004).

\bibitem {Gabrielse2003}
B. D'Urso, B. Odom, and G. Gabrielse, Phys. Rev. Lett. \textbf{90}, 043001 (2003).

\bibitem {Rempe2002}
T. Fischer, P. Maunz, P. W. H. Pinkse, T. Puppe, and G. Rempe, Phys. Rev.
Lett. \textbf{88}, 163002 (2002).

\bibitem {Morrow2002}
N. V. Morrow, S. K. Dutta, and G. Reithel, Phys. Rev. Lett. \textbf{88},
093003 (2002).

\bibitem {Geremia2004}
J.M. Geremia, J. K. Stockton, H. Mabuchi, Science \textbf{304}, 270
(2004); D. Oblak, P. G. Petrov, C. L. Garrido Alzar, W. Tittel, A. K.
Vershovski, J. K. Mikkelsen, J. L. S\o rensen, and E. S. Polzik, Phys.
Rev. A \textbf{71}, 43807 (2005).

\bibitem {Wiseman1993}
H.M. Wiseman and G. J. Milburn, Phys. Rev. Lett. \textbf{70}, 548-551
(1993).

\bibitem {Vitali2002}
S. Mancini, D. Vitali, and P. Tombesi , Phys. Rev. Lett. \textbf{80},
688-691 (1998); D. Vitali, S. Mancini, L. Ribichini, and P. Tombesi,
Phys. Rev. A \textbf{65}, 063803 (2002).

\bibitem {Milatz1953}J. M. W. Milatz and J. J. Van Zolingen, Physica
\textbf{XIX}, 181(1953).

\bibitem {Cohadon1999}
P. F. Cohadon, A. Heidmann, and M. Pinard, Phys. Rev. Lett. \textbf{83},
3174 (1999).

\bibitem {Dorner2002}
U. Dorner and P. Zoller Phys. Rev. A \textbf{66}, 023816 (2002).

\bibitem {Bushev2004}
P. Bushev, A. Wilson, J. Eschner, C. Raab, F. Schmidt-Kaler, C. Becher,
and R. Blatt, Phys. Rev. Lett. \textbf{92}, 223602 (2004);
J. Eschner, Ch. Raab, F. Schmidt-Kaler, and R. Blatt, Nature
\textbf{413}, 495-498 (2001).

\bibitem {GZ}C. W. Gardiner and P. Zoller, \textit{Quantum Noise} (Springer,
Berlin, 2004), and references cited.

\bibitem {Buchler1999}
B. C. Buchler, M. B. Gray, D. A. Shaddock, T. C. Ralph, D. E. McClelland, Opt.
Lett. \textbf{24}, 259 (1999).

\bibitem {Masalov1994}
A. V. Masalov, A. A. Putilin, and M. V. Vasilyev, J. Mod. Optics \textbf{41},
1941 (1994).

\bibitem {Haus1986}
H. A. Haus, Y. Yamamoto, Phys. Rev. A \textbf{34}, 270-292 (1986).

\bibitem {Shapiro1987}
J. H. Shapiro, G. Saplakoglu, S.-T. Ho, P. Kumar, B. E. A. Saleh,
M. C. Teich, J. Opt. Soc. Am. B \textbf{4}, 1604-1620 (1987).

\bibitem {Steixner2005}
V. Steixner, P. Rabl, and P. Zoller, accepted by Phys. Rev. A,
quant-ph/0506187.

\end{thebibliography}
\end{document}